\documentclass[aps,amsmath,amssymb,prb,superscriptaddress,twocolumn,letterpaper,10pt]{revtex4-1}
\usepackage{graphicx}
\usepackage{txfonts}
\usepackage{bm}
\usepackage{mathrsfs}
\usepackage{mathtools}
\mathtoolsset{showonlyrefs}
\usepackage{comment}
\usepackage{array}
\usepackage{appendix}
\usepackage{color}
\newcolumntype{C}[1]{>{\centering\arraybackslash}p{#1}}

\newcommand{\al}{\alpha}

\newcommand{\g}{\gamma}
\newcommand{\de}{\delta}

\newcommand{\p}{\pi}

\newcommand{\f}{\phi}
\newcommand{\x}{\chi}
\newcommand{\w}{\omega}
\newcommand{\W}{\Omega}
\newcommand{\De}{\Delta}
\newcommand{\G}{\Gamma}

\newcommand{\pd}{\partial}

\newcommand{\round}[1]{\left( #1 \right)}
\renewcommand{\square}[1]{\left[ #1 \right]}

\newcommand{\mat}[4]{\left(\begin{array}{cc}#1&#2\\#3&#4\end{array}\right)}

\newcommand{\ang}[1]{\left\langle #1 \right\rangle}

\newcommand{\bra}[1]{\left\langle #1\right|}
\newcommand{\ket}[1]{\left| #1\right\rangle}

\newcommand{\beq}{\begin{equation}}
\newcommand{\eeq}{\end{equation}}
\newcommand{\Beq}{\begin{eqnarray}}
\newcommand{\Eeq}{\end{eqnarray}}
\newcommand{\bml}{\begin{multline}}

\newcommand{\bea}{\begin{align}}
\newcommand{\ena}{\end{align}}
\newcommand{\bsp}{\begin{split}}
\newcommand{\esp}{\end{split}}
\newcommand{\down}{\downarrow}
\newcommand{\up}{\uparrow}

\newcommand{\bM}{{\boldsymbol M}}

\newcommand{\ex}{\hat{\boldsymbol x}}

\newcommand{\ey}{\hat{\boldsymbol y}}
\newcommand{\ez}{\hat{\boldsymbol z}}

\newcommand{\bj}{{\boldsymbol j}}

\newcommand{\bE}{{\boldsymbol E}}

\newcommand{\bB}{{\boldsymbol B}}

\newcommand{\hsig}{\hat{\sigma}}

\newcommand{\sA}{\mathscr{A}}

\newcommand{\hH}{\hat{H}}

\newcommand{\hphi}{\hat{\varphi}}

\newcommand{\bn}{\boldsymbol{n}}

\newcommand{\bve}{{\boldsymbol{\varepsilon}}}

\newcommand{\tal}{\tilde{\al}}

\newcommand{\tih}{\tilde{h}}

\begin{document}
\title{Spin superfluid Josephson quantum devices}
\author{So Takei}
\affiliation{Department of Physics, Queens College of the City University of New York, Queens, NY 11367, USA}
\author{Yaroslav Tserkovnyak}
\affiliation{Department of Physics and Astronomy, University of California, Los Angeles, CA 90095, USA}
\author{Masoud Mohseni}
\affiliation{Google Research, Venice, CA 90291, USA}
\date{\today}

\begin{abstract}
A macroscopic spintronic qubit based on spin superfluidity and spin Hall phenomena is proposed. This magnetic quantum information processing device realizes the spin-supercurrent analog of the superconducting phase qubit, and allows for full electrical control and readout. We also show that an array of interacting magnetic phase qubits can realize a quantum annealer. These devices can be built through standard solid-state fabrication technology, allowing for scalability. However, the upper bound for the operational temperature can, in principle, be higher than the superconducting counterpart, as it is ultimately governed by the magnetic ordering temperatures, which could be much higher than the critical temperatures of the conventional superconducting devices.
\end{abstract}
\maketitle
\newpage

\section{Introduction}
Macroscopic quantum phenomena in magnetic systems has been a topic of active research for a number of years. Manifestations of such phenomena have been discussed in the context of ferromagnetic domain walls~\cite{chudnovskyPRB92,*tataraPRL94,*braunPRB96,*brookeNAT01}, magnetic nanoparticles~\cite{barbaraPLA90,chudnovskyPRL88,*awschalomSCI92,*giderSCI95} as well as molecular magnets~\cite{chioleroPRL98,*meierPRL01,*barbaraJMMM99,*wernsdorferSCI99}. Molecular magnets, in particular, have garnered much attention for their potential utility in quantum information technology~\cite{leuenbergerNAT01}. Despite these activities, research addressing the possibility of macroscopic qubits in magnetic systems still remains essentially absent. In this work, we propose the first macroscopic spin-based qubit by combining two recent advancements in the field of spintronics: spin superfluidity and spin Hall phenomena. 

Spin superfluidity explores how analogs of conventional superfluidity can be realized in magnetically ordered systems~\cite{soninAP10}. The Josephson effect in conventional superfluids involves a dissipationless mass flow between two weakly-coupled superfluids, and relies on macroscopic phase coherence of the superfluids each characterized by a U(1) order parameter. Magnetic order in certain insulating magnets are also described by macroscopic U(1) order parameters~\cite{soninAP10}. In analogy with conventional superfluidity, coupling two such magnets gives rise to the {\em magnetic} Josephson effect involving dissipationless (superfluid) flow of spin angular momentum between the magnets~\cite{schillingANP12}. Spin Hall phenomenology is often discussed in the context of a bilayer system consisting of a normal metal with strong spin-orbit coupling and an insulating ferromagnet~\cite{brataasCHA12}. The combination of spin-orbit coupling in the metal and exchange coupling (between the conduction electron spins and the magnetic moments in the insulator) at the metal$|$magnet interface allow angular momentum to be transferred from the metal's crystal lattice to the ferromagnet, thus engendering macroscopic coupling between the electrical current in the metal and the magnetic order parameter~\cite{kajiwaraNAT10,*sandwegPRL11,*burrowesAPL12,*hahnPRB13}.

In this work, we show that a heterostructure consisting of magnetic insulators and a normal metal with strong spin-orbit coupling realizes a magnetic analog of the superconducting phase qubit~\cite{martinisBOOK09} and permits full electrical control and readout via spin Hall phenomena (see Fig.~\ref{setup}). We refer to this device as the {\em magnetic phase qubit}. While electrical readout of molecular qubits can be challenging~\cite{leuenbergerNAT01}, spin Hall phenomena offers an alternative, and a possibly more straightforward, method for electrical control and readout of the magnetic qubit. The device is an example of a macroscopic qubit that can be constructed from solid state materials and so should offer the same advantages as the superconducting qubits of strong inter-qubit coupling and scalability. The magnetic qubit is based on antiferromagnets, which are known to display macroscopic quantum behavior at higher temperatures than ferromagnets~\cite{barbaraPLA90}. We estimate the operational temperature of the qubit to be $T\sim 3~K$, which, in contrast to superconducting qubits, can be achieved by a pure helium-4 cryostat without a dilution fridge. Furthermore, the zero net magnetization in antiferromagnets implies that they do not generate stray fields; this is advantageous when coupling qubits, as it will eliminate any unwanted magnetic cross-talk between neighboring qubits. We find that the the relaxation time scale for our qubit is relatively small, with $T_1\sim10$~ns, because it is set in our device by the internal magnetic dynamics of the antiferromagnet (known to be in the THz range)~\cite{kefferPR52,*gomonayLTP14}. Finally, we show that a coupled array of these qubits can realize a quantum annealer~\cite{kadowakiPRE98}, which could be used to solve certain hard optimization problems and machine learning tasks.

\section{Model and theory}
\begin{figure}[t]
\centering
\includegraphics*[scale=0.38]{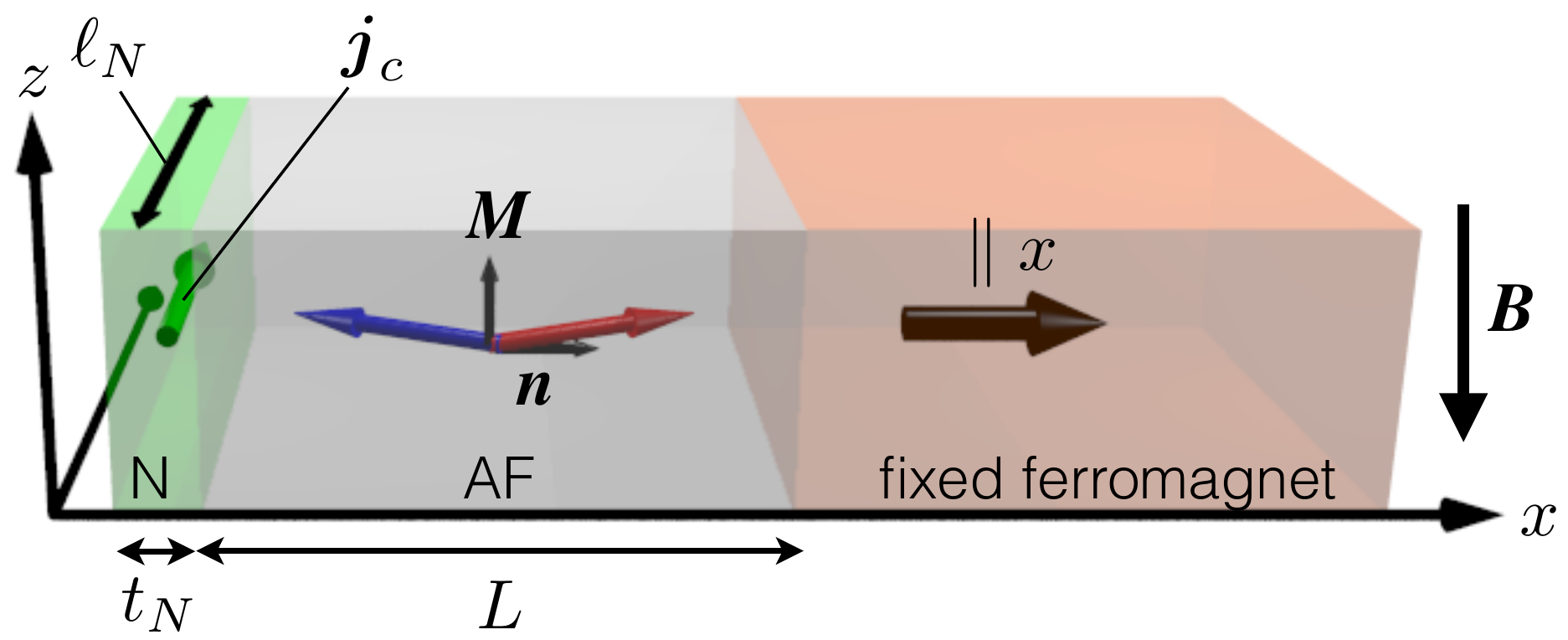}
\caption{The magnetic phase qubit. A mono-domain antiferromagnet (AF) is exchange-coupled to a fixed mono-domain ferromagnet (with its macrospin oriented along the $x$ axis) and attached to a normal metal (N) with strong spin-orbit coupling. Charge current $j_c$ is applied in the $y$ direction in the N. Here, $\bM$ denotes the total spin angular momentum in the AF while $\bn$ denotes the unit vector parallel to the N\'eel order.}
\label{setup}
\end{figure}

The prototype magnetic phase qubit consists of an isotropic (Heisenberg) antiferromagnetic insulator (AF) attached to a normal metal (N) with strong spin-orbit coupling, exchange-coupled to a fixed ferromagnetic layer, and subjected to an external magnetic field in the $-z$ direction (see Fig.~\ref{setup}). We consider low enough temperatures such that only the lowest mode is excited in the AF and it can be considered in the mono-domain limit (upper bound on temperature for mono-domain operation is estimated in Sec.~\ref{dis}). A bipartite AF can be characterized by two variables, $\bn(t)$ and $\bM(t)$, representing the N\'eel vector and the total spin angular momentum in the AF, respectively~\cite{takeiPRB14}. These variables are chosen to satisfy $|\bn|=1$ and $\bn\cdot\bM=0$, the presence of strong N\'eel order implying $|\bM|/S\ll1$ ($S$ being twice the total spin angular momentum on one sublattice in units of $\hbar$). The AF Lagrangian and its coupling to the ferromagnet can then be written as $L=\hbar\dot\bn\cdot(\bM\times\bn)-\bM^2/2\x+bM_z+Jn_x$, where $b = \hbar\g_0|\bB|$ (with $\g_0$ the gyromagnetic ratio and $\bB$ the magnetic field), $J$ is the AF-ferromagnet exchange coupling, and $\x$ is the AF spin susceptibility. The N is taken to be a metal with strong spin-orbit coupling [e.g., platinum (Pt)], and is assumed to be a metallic film parallel to the $yz$ plane obeying Ohm's law $\rho\bj_c=\bE$, where $\bE$ is the electric field, $\bj_c$ is the linear charge current density and $\rho$ is its 2D resistivity. 

The current in the N can be used to manipulate the spins in the AF. This is achieved by a combination of spin-orbit coupling in the N and the exchange coupling at the N$|$AF interface, which allow angular momentum to be transferred from the N crystal lattice to the AF spins, effectively generating a macroscopic coupling between the current and the AF variables $\bn$ and $\bM$~\cite{hahnEPL14,*wangPRL14af,*moriyamaAPL15}. Through this magnetoelectric coupling, the current generates torques on the AF spins, thus resulting in an injection of spin current $\bj_s$ into the AF, and AF dynamics, in turn, induces an electromotive force $\bve$ in the N. Spin Hall phenomenology allows one to write down general expressions for $\bve$ and $\bj_s$ that respect certain crystalline and structural symmetries at the interface~\cite{tserkovnyakPRB14}. In the presence of full translational and rotational symmetries in the $yz$ plane and the breaking of reflection symmetry along the $x$ axis, $\bve$ within spin Hall phenomenology reads
\beq
\label{emf}
\bve=\vartheta(\bn\times\dot\bn)\times\ex\ ,
\eeq
and the spin current entering the AF takes the form
\beq
\label{torque}
\bj_s=\sA\vartheta\bn\times(\ex\times\bj_c)\times\bn\, .
\eeq
Here, $\sA$ is the cross-sectional area of the interface and $\vartheta$ is a phenomenological torque coefficient for the N$|$AF interface. The fact that the same coefficient $\vartheta$ enters both Eqs.~\eqref{emf} and \eqref{torque} is a consequence of Onsager reciprocity~\cite{brataasCHA12}. In Eq.~(\ref{emf}), we have ignored the so-called reactive contribution $\propto \dot\bn\times\ex$ (also allowed by symmetry), as it typically gives a significantly smaller contribution compared to the term in Eq.~\eqref{emf} for diffusive metals like the N considered here~\cite{tserkovnyakPRB14}. Eq.~\eqref{emf} was obtained under the assumption that the AF N\'eel order is collinear; this is a good approximation as long as all energy scales are small with respect to the AF exchange scale.

The analysis of the AF dynamics here follows Ref.~\onlinecite{takeiPRB14} very closely. For $\bj_c=0$, $\bn$ lies in the $xy$ plane and the field gives a finite total spin $\bM=\x b\ez$. For a nonzero $\bj_c=j_c\ey$, the AF dynamics can be described in terms of a dynamic N\'eel vector confined within the $xy$ plane, i.e., $\bn=(\cos\varphi(t),\sin\varphi(t),0)$, and an additional canting $\xi(t)$ defined via $\bM(t)=[\x b+\xi(t)]\ez$. Linearizing the AF dynamics in $\xi(t)$ and $\dot\varphi(t)$ and phenomenologically introducing viscous magnetic damping in the AF, the coupled dynamics of the current and the AF variables become~\cite{takeiPRB14}
\beq
\label{afdyn}
\hbar\dot\varphi=\frac{\xi}{\x}+\frac{\hbar\al'}{S}\dot\xi,\qquad \hbar\dot\xi=-J\sin\varphi+j-\hbar\al S\dot\varphi\ ,
\eeq
and $\rho\bj_c=\bE+\bve$, where $j\equiv\sA\vartheta j_c$, and $\al$, $\al'$ are two independent Gilbert damping parameters. The N subjects the AF to additional viscous damping, which modifies the spin current term in Eq.~\eqref{torque} to $\bj_s\rightarrow\bj'_s\equiv\bj_s-\hbar\al^{\up\down}S\bn\times\dot\bn$. If N is a perfect spin sink (as for Pt), $\al^{\up\down}=\sA g^{\up\down}/4\p S$, where $g^{\up\down}$ is the (real part of the) effective interfacial spin-mixing conductance per unit area. This shifts $\al\rightarrow\tal=\al+\al^{\up\down}$ and $\al'\rightarrow\tal'$ in Eq.~\eqref{afdyn} (the precise expression for $\tal'$ is unimportant here since it will not play a crucial role in the following discussion). As Eq.~\eqref{afdyn} neglects the mode corresponding to the N\'eel vector oscillating out of the $xy$ plane (with a gap of $b$), it is an appropriate description of the AF dynamics as long as $b>\sqrt{J/\chi}$. 

For $\tal=\tal'=0$, noticing that $\xi$ is the momentum conjugate to $\varphi$, the effective Hamiltonian corresponding to Eq.~\eqref{afdyn} is given by ($\eta\equiv j/J$)
\beq
\label{h}
H_{\rm eff}=\frac{\xi^2}{2\x}-J(\cos\varphi+\eta\varphi)\equiv\frac{\xi^2}{2\x}+U(\varphi)\ .
\eeq
Well below the N\'eel temperature, $\x\sim N/J_{\rm AF}$, where $N$ is the total number of spins in the AF and $J_{\rm AF}$ is its internal AF exchange scale. Since $N\gg1$, it is reasonable to assume the regime $J\x\gg1$. Then the Josephson term (proportional to $J$) dominates in $H_{\rm eff}$, which then has the same form as the standard current-biased Josephson junction with the charging energy given by $\x^{-1}$ and the Josephson energy by the exchange coupling $J$. Here, the ``bias current" $j$ is a flow of ($z$-polarized) spin across the N$|$AF interface, and is fully controllable using the external {\em charge} current $j_c$. Interestingly, while the torque in Eq.~\eqref{torque} [which gives rise to the term proportional to $\eta$ in Eq.~\eqref{h}] is dissipative in the sense that it is odd under time-reversal, its effects enter the AF dynamics in a way still accountable by purely Hamiltonian dynamics. 

\section{Defining the Qubit}
The following results closely follow those for the superconducting phase qubits~\cite{zagoskinPC07,martinisPRB03}. We emphasize that the qubit is operated at $\eta\lesssim 1$ to ensure only a few states are present in a washboard potential minimum. In this regime, the potential about $\varphi=\p/2$ can be approximated by a cubic form $U(\f)\approx (J-j)\f-J\f^3/6+{\rm const.}$ (with $\f\equiv\varphi-\p/2$). A local minimum is located at $\f_0=-\sqrt{2(1-\eta)}$, and the plasma frequency corresponding to the quadratic curvature at the minimum is given by
\beq
\label{hwp}
\hbar\w_p=\sqrt{J/\x}\square{2\round{1-\eta}}^{1/4}.
\eeq
Imposing the canonical commutation relation $[\hphi,\hat\xi]=i$ leads to the quantization of energy levels inside the cubic potential, and when $\eta\lesssim1$  only a few quantum states are bound in each of the local minima. We label the lowest three energy eigenvalues by $E_0$, $E_1$ and $E_2$, and the corresponding states by $\ket{0}$, $\ket{1}$ and $\ket{2}$. Taking account of the cubic anharmonicity to second-order in perturbation theory, the separation between the two lowest pairs of energy levels become $E_1-E_0\equiv\hbar\w_{10}\approx\hbar\w_p\round{1-r}$ and $E_2-E_1\equiv\hbar\w_{21}\approx\hbar\w_p\round{1-2r}$, where the deviations from the harmonic limit is quantified in terms of $r\equiv(5/36)(\hbar\w_p/\De U)$, $\De U\equiv J(4\sqrt{2}/3)(1-\eta)^{3/2}$ being the energy barrier for the particle to escape from a local minimum. 

\subsection{Qubit control}
The state of the qubit can be controlled by introducing an ac component to $j$ oscillating at frequency $\w_{10}$ [$j(t)=j_{\rm dc}+j_{\rm ac}(t)\cos(\w_{10}t)$], where $j_{\rm dc}$ is the dc component and $j_{\rm ac}(t)$ modulates the amplitude of the ac component. The ac component to $j$ can be introduced simply by adding an ac component to the charge current. To inhibit transitions between states $\ket{1}$ and $\ket{2}$, one requires the temporal variations of $j_{\rm ac}(t)$ to be slow compared to $2\p/(\w_{21}-\w_{10})$. In this limit, and for temperatures $T\ll\hbar\w_{10}/k_B$, the Hilbert space for the qubit is spanned by the two lowest states $\ket{0}$ and $\ket{1}$, and the effective Hamiltonian reads 
\beq
\label{heff}
\hH_{\rm eff}=-\frac{\hbar\w_{10}}{2}\hsig^z-\frac{j_{\rm ac}(t)\g}{2}\mat{0}{e^{i\w_{10}t}}{e^{-i\w_{10}t}}{0}\ ,
\eeq
where $\g\equiv\sqrt{1/2\x\hbar\w_{10}}$, and we have invoked the rotating wave approximation and the off-diagonal elements were computed using harmonic oscillator states due to the small non-linearity of the system. The qubit can undergo Rabi oscillations, where the probability for transition between $\ket{0}$ and $\ket{1}$ states $P_{0\rightarrow1}(t)=\sin^2[\W_{\rm ac}(t)t]$, where $\hbar\W_{\rm ac}(t)\equiv j_{\rm ac}(t)\g/2$. The application of an ac pulse over time interval $\De t=\p/2\W_{\rm ac}\equiv t_0$ drives the qubit from state $\ket{0}$ to $\ket{1}$. With a pulse of length $\De t=t_0/2$, the qubit can be manipulated into the superposition state $(\ket{0}+\ket{1})/\sqrt{2}$.

\subsection{Qubit readout}
The state of the qubit can be read out by adiabatically lowering the potential barrier $\De U$ close to $E_1$ such that the qubit in state $\ket{1}$ is (exponentially) more likely to tunnel out of a local minimum than that in state $\ket{0}$. The limit $\De U\rightarrow E_1$ is achieved as $\eta\rightarrow1$~[see Sec.~\ref{dis} for further details]. Once the ``particle" tunnels out of the local minimum, it will descend down the washboard potential generating a finite $\langle\dot\varphi\rangle$, which induces an electromotive force $\bve\approx\vartheta\langle\dot\varphi\rangle\ey$, so that Ohm's law in the N is modified to $(\rho-\de\rho)j_c=E$. Here, we neglect the $z$ component of $\bve$ as it is counteracted by an electrostatic buildup along the $z$ axis as long as the magnetic dynamics are slow compared to the relevant RC time of the normal metal. We can estimate $\de\rho$ by considering the motion of the particle classically. Approximating the washboard potential as a downward-sloping line connecting all of the local minima, the descending particle (damped by Gilbert damping $\tal$) will reach a terminal velocity $\dot\varphi_t=\sA\vartheta j_c/\hbar\tal S$. The correction to the effective resistivity thus reads $\de\rho=\sA\vartheta^2/\hbar\tal S$. For a fixed $j_c$, $\de\rho$ gives rise to a decrease in voltage of $\de V= \de\rho j_c\ell_N$ across the N, where $\ell_N$ is the length of the N in the $y$ direction. The tunneling out of state $\ket{1}$ can be detected by detecting this voltage decrease. To detect state $\ket{0}$, one may transfer the qubit to state $\ket{1}$ using a resonant ac pulse, and then detect state $\ket{1}$.

\section{Decoherence}
\subsection{Decoherence due to Gilbert damping}
Decoherence in the magnetic phase qubit arises from various environmental degrees of freedom that couple to the macrospin, e.g., the phonons in the AF as well as the electron continuum in the N. Dissipation due to these environments have been accounted for by Gilbert damping. A damped macrospin experiences a stochastic force required to exist by the fluctuation-dissipation theorem~\cite{landauBOOKv5}. The stochastic force can be described by introducing a random component, $\de H_{\rm eff}(t)=h_\varphi(t)\varphi+h_{\xi}(t)\xi$ to $H_{\rm eff}$, where $h_{\varphi}(t)$ and $h_{\xi}(t)$ are the stochastic fields. These stochastic fields modify Eq.~\eqref{afdyn} to
\beq
\begin{aligned}
\label{dynr}
\hbar\dot\xi&=-J\sin\varphi+j-\hbar\tal S\dot\varphi-h_\varphi\ , \\
\hbar \dot\varphi&=\frac{\xi}{\x}+\frac{\hbar\tal'}{S}\dot\xi+h_{\xi}\ .
\end{aligned}
\eeq
Accounting for both thermal and quantum fluctuations, the symmetrized correlation functions for the stochastic fields read~\cite{landauBOOKv5,martinisPRB03}
\beq
\begin{aligned}
\label{corrs}
\frac{1}{2}\langle h_\varphi(t)h_\varphi(0)+h_\varphi(0)h_\varphi(t)\rangle&=\int\frac{d\w}{2\p}\ \x_\varphi(\w)e^{-i\w t}\ , \\
\frac{1}{2}\langle h_\xi(t)h_\xi(0)+h_\xi(0)h_\xi(t)\rangle&=\int\frac{d\w}{2\p}\ \x_\xi(\w)e^{-i\w t}\ ,
\end{aligned}
\eeq
where $\x_\varphi(\w)=\hbar^2\tal\w S\coth(\hbar\w/2k_BT)$ and $\x_\xi(\w)=(\hbar^2\tal'\w/S)\coth(\hbar\w/2k_BT)$. We now promote both $\varphi$ and $\xi$ to quantum-mechanical operators, and project $\de H_{\rm eff}(t)$ to the two lowest quantum states $\ket{0}$ and $\ket{1}$. If the states are again approximated as harmonic oscillator states, we have 
\beq
\begin{aligned}
\bra{0}\hat\varphi\ket{1}\approx\g,\quad\bra{0}\hat\xi\ket{1}\approx(2i\g)^{-1}\ ,
\end{aligned}
\eeq
and the diagonal elements vanish. The stochastic contribution to the two-level Hamiltonian then becomes $\de\hH_{\rm eff}(t)=\tih_x(t)\hsig^x+\tih_y(t)\hsig^y$, with $\tih_x(t)=\g h_\varphi(t)$ and $\tih_y(t)=(2\g)^{-1}h_{\xi}(t)$. Here, the frequency spectra of the new noise fields obey $\tilde\x_x(\w)=\g^2\x_\varphi(\w)$ and $\tilde\x_y(\w)=(\tal'/\tal)(2\g^2 S)^{-2}\x_\varphi(\w)$.

To estimate the $T_1$ and $T_2$ times, we consider the undriven qubit, i.e., $j_{\rm ac}(t)=0$. We have $2\g^2 S\sim N(J\x)^{-1/2}\gg1$, so we may ignore the fluctuations $\tih_y(t)$ in this estimate. The relevant two-level Hamiltonian then reads
\beq
\label{heffs}
\hH_{\rm eff}\approx-\frac{\hbar\w_{10}}{2}\hsig_z+\tih_x(t)\hsig_x\ .
\eeq
The noise term causes transitions between the two qubit states. If the qubit starts in state $\ket{1}$ at $t=0$, the amplitude for the qubit to be in the ground state $\ket{0}$ at time $t$ is given through Fermi's golden rule by
\beq
c_0(t)\approx\frac{1}{i\hbar}\int_0^tdt'\bra{0}\tih_x(t)\hsig_x\ket{1}e^{i\w_{10}t'}\ .
\eeq
Similarly, if the qubit starts in state $\ket{0}$ at $t=0$, the amplitude for the qubit to be in the excited state $\ket{1}$ at time $t$ is given by
\beq
c_1(t)\approx\frac{1}{i\hbar}\int_0^tdt'\bra{1}\tih_x(t)\hsig_x\ket{0}e^{-i\w_{10}t'}\ .
\eeq
The longitudinal relaxation rate is given by
\beq
\G_1=\dot p_0(t)+\dot p_1(t)\ ,
\eeq
where $p_0(t)\equiv\langle|c_0(t)|^2\rangle$ and $p_1(t)\equiv\langle|c_1(t)|^2\rangle$. We find
\beq
\begin{aligned}
\label{c0}
p_0(t)+p_1(t)=\frac{\g^2}{\hbar^2}\int\frac{d\w}{2\p}\frac{\sin^2\round{\frac{\w-\w_{10}}{2}t}}{\round{\frac{\w-\w_{10}}{2}}^2}\x_\varphi(\w)\ .
\end{aligned}
\eeq
Most of the integral contribution in Eq.~\eqref{c0} comes from $\w\sim\w_{10}$, where $\x(\w)$ is a slow-varying function of $\w$ but its pre-factor inside the integrand is strongly peaked. Then the integral can be approximated by evaluating $\x_\varphi(\w)$ at $\w=\w_{10}$ and taking it out of the integral. This then leads to 
\beq
\G_1\approx\frac{\g^2\x_\varphi(\w_{10})}{\hbar^2}=\frac{\tal S}{\x\hbar}\ ,
\eeq
where the result holds in the low temperature regime, $k_BT\ll\hbar\w_{10}$. Since the stochastic fields only involve transverse components, $\G_2=\G_1/2$.

\begin{figure}[t]
\centering
\includegraphics*[scale=0.33]{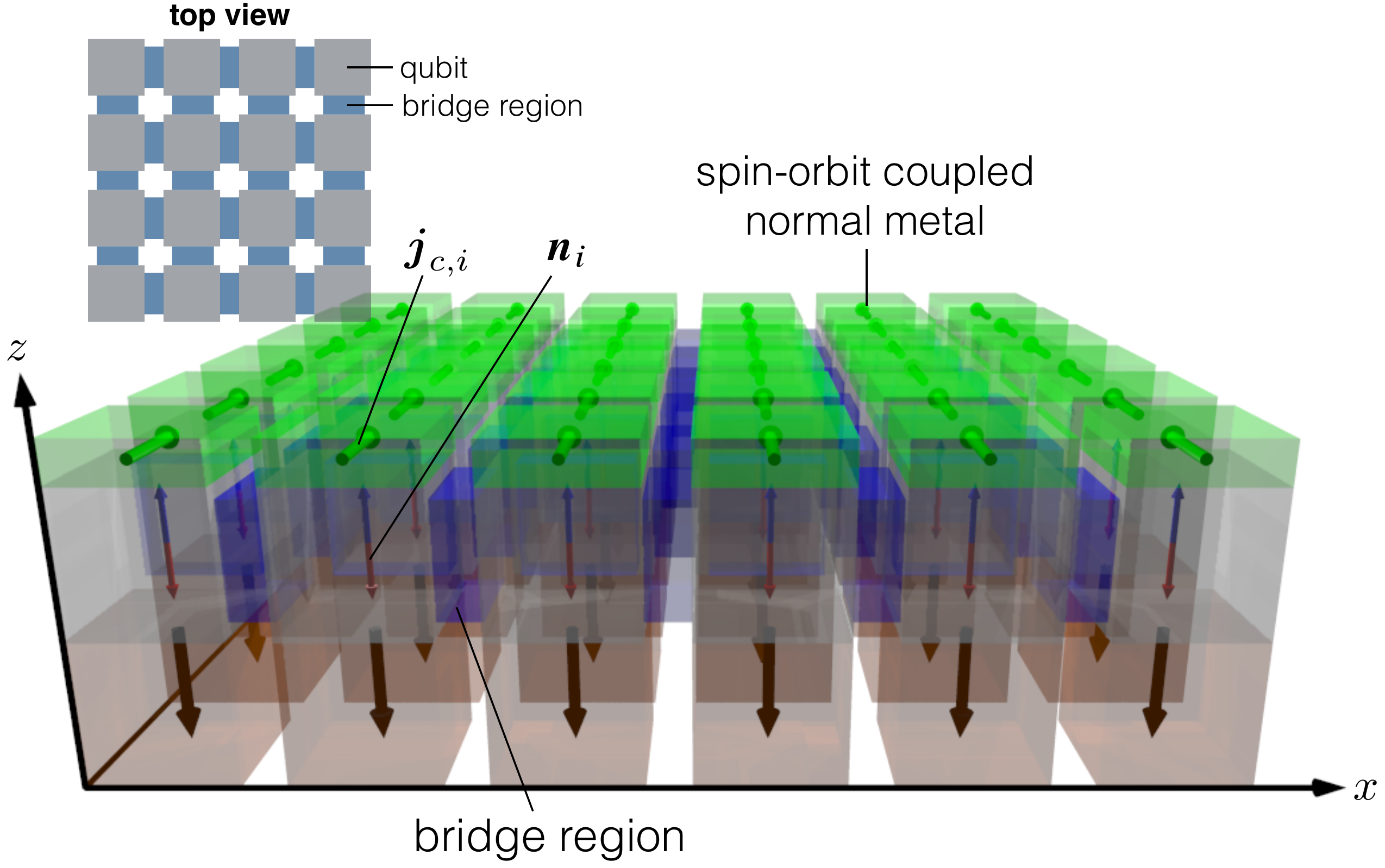}
\caption{A coupled array of magnetic phase qubits. Each qubit is coupled to its own normal metal with independently tunable dc and ac currents. The ferromagnetic region (shaded in blue) mediate a coupling between nearest neighbor AFs and give rise to inter-qubit coupling.}
\label{fig2}
\end{figure}

\subsection{Decoherence due to current noise}
Decoherence can also arise from fluctuations in the electrical current used to control the qubit. Here, we consider decoherence due to fluctuations in the dc component of the charge current $j_{\rm c}$ and for $j_{\rm ac}=0$. In the presence of the fluctuations, the effective two-level Hamiltonian Eq.~\eqref{heff} is endowed by a stochastic contribution given by
\beq
\hH_0(t)=-\frac{\hbar\w_{10}}{2}\hsig_z-\frac{\hbar}{2L}\frac{\pd\w_{10}}{\pd j_c}\de I_c(t)\hsig_z\ .
\eeq
Here, $\de I_c(t)$ is the noise in the dc charge current driven through the N. Defining the symmetrized noise correlators for the stochastic fields as
\beq
\label{snc}
\ang{\de I_c(t)\de I_c(0)+\de I_c(0)\de I_c(t)}=\int\frac{d\w}{\p} S_I(\w)e^{-i\w t}\\
\eeq
and using standard results for noise and decoherence in quantum two-level systems, charge current noise gives rise to pure dephasing rate of
\beq
\label{dr}
\G_\varphi=\frac{1}{4L^2}\round{\frac{\pd\w_{10}}{\pd j_c}}^2S_I(0)\ .
\eeq
\section{Qubit coupling}
One may consider coupling many magnetic phase qubits in an array as shown in Fig.~\ref{fig2}. Every qubit is coupled to its own N so that its splitting $\hbar\w_{10,i}$ and the Rabi frequency $\W_{{\rm ac},i}$ can be tuned independently by adjusting the dc and ac amplitudes of $\bj_{c,i}(t)$, respectively ($i$ here labels the qubits). We connect neighboring AFs via a metallic ``bridge" (shaded in blue) that gives rise to an effective exchange interaction $J'_{ij}$ between the $i$-th and $j$-th N\'eel vectors. This coupling may be generated if the AFs expose more spins from one sublattice than the other to the metallic link, and if the conduction electrons in the metallic link (with thickness $d$) mediate an RKKY interaction between these net spin moments. The RKKY interaction is generally an effective exchange interaction between two spin moments mediated by the conduction electrons in the metallic link. Using a standard result, the effective RKKY coupling (for $k_Fd\gg1$) is $\propto J_0^2\cos(k_Fd)/(k_Fd)^3$, where $k_F$ is the Fermi wave-vector of the metallic spacer and $J_0$ is the interfacial exchange coupling between the spin densities in the link and the AF~\cite{brunoBOOK93}. Therefore, electrostatic gating of each individual metallic spacer allows one to control the electron density (and thus $k_F$) of each spacer individually and allow local dynamic control of both the sign and the magnitude of $J'_{ij}$. 

This coupling between the neighboring AFs gives rise to the interaction of the form $H_{\rm int}=-\sum_{\langle i<j\rangle}J'_{ij}\cos(\varphi_i-\varphi_j)$, where $\langle i,j\rangle$ label the nearest neighbor qubits. Quantizing this Hamiltonian and projecting it to the two logical states of each qubit, the interaction Hamiltonian reads
\beq
\label{hint}
\hH_{\rm int}=\frac{1}{4}\sum_{\langle i<j\rangle}J'_{ij}\g^2\hsig^x_i\hsig^x_j\ .
\eeq
Here, we have neglected terms of order $J'/J$ that only renormalize the qubit frequency $\w_{10}$. The coupled array of qubits is then described by the total Hamiltonian $\hH_t=\hH_0+\hH_{\rm int}$, where
\beq
\hH_0=-\sum_i\square{\frac{\hbar\w_{10,i}}{2}\hsig^z_i+\frac{j_{{\rm ac},i}\g}{2}\mat{0}{e^{i\w_{10,i}t}}{e^{-i\w_{10,i}t}}{0}}\ .
\eeq

The coupled array of magnetic phase qubits as shown in Fig.~\ref{fig2} can be used to realize a quantum annealer. A quantum annealer solves hard optimization problems by finding the ground state of a ``problem Hamiltonian" through a process involving quantum fluctuations. It can be implemented with a classical Ising Hamiltonian that encodes the computational problem, and a transverse field term, $\hH_{\rm QA}(t)=-A(t)\sum_i\hsig_i^x+B(t)\hH_{\rm Ising}$, where $\hH_{\rm Ising}=-\sum_ih_i\hsig_i^z-\sum_{i<j}J_{ij}\hsig_i^z\hsig_j^z$, $i,j$ label lattice sites, and $t$ runs from $0$ to $t_f$. At $t=0$ and temperature $T$, the quantum annealing process begins in the limit of strong transverse field, i.e., $A(0)\gg \{k_BT,|B(0)h_i|,|B(0)J_{ij}|\}\ \ \forall i,j$, when the quantum-mechanical fluctuations dominate, and then one gradually decreases $A(t)$ and increases $B(t)$ such that the state of system approaches a classical bit string that ideally representing the ground state of the problem Hamiltonian. Generally, at sufficiently low temperatures, the quality of solution is improved when time-scale of annealing schedule is substantially larger than the inverse of the square of the minimum gap between instantaneous ground-state and the first existed state of the effective many-body Hamiltonian given by $\hH_{\rm QA}$. Rotating the spin axes of all the qubits by $\pi/2$ about the $y$ axis (such that $\hsig_{z,x}\rightarrow\pm\hsig_{x,z}$), $\hH_t$ can be shown to mimic the Hamiltonian $\hH_{\rm QA}$ so the coupled array of magnetic phase qubits above can be used to implement quantum annealing. 

\section{Discussion}
\label{dis}
The qubit must operate at temperatures $k_BT\ll \hbar\w_{10}\equiv k_BT_+\sim\sqrt{J\x^{-1}}$. A good candidate material for a Heisenberg AF is a cubic perovskite KNiF$_3$~\cite{jonghAIP74}, a spin-1 AF with $J_{\rm AF}\sim 0.01$~eV and lattice constant of $a\approx 4\AA$. Let us consider a cube-shaped KNiF$_3$ with side length $L=50$~nm. Assuming that the exchange coupling between the AF and the ferromagnet arises from the exchange bias effect with a corresponding field of $B_{\rm ex}\sim100$~mT~\cite{meiklejohnPR57,*noguesJMMM99}, we may take $J\sim \hbar\g B_{\rm ex}(L/a)^3$. Since $\x\sim (L/a)^3/J_{\rm AF}$, we obtain $T_+\approx3~{\rm K}$. Using $\al\sim10^{-5}$, and assuming that the damping enhancement $\al^{\up\down}$ due to the N just exceeds $\al$, we take $\tal\sim10^{-5}$ and we obtain $T_1\sim10$~ns. The non-linearity necessary for the qubit to remain within the two-level subspace requires $\eta\sim1$. The (linear) current density necessary to achieve this is $\bar j_c\sim J/\sA\vartheta$. Here, $\vartheta=\hbar\tan\theta_{\rm SH}/2et_N$, where $\theta_{\rm SH}$ is the effective spin Hall angle for the N$|$AF interface and $t_N$ is the N thickness~\cite{tserkovnyakPRB14}. Using $\theta_{\rm SH}\sim0.1$ (appropriate for a Pt contact), $t_N=10$~nm, and setting all other parameters to the values above, we obtain $\bar j_c\sim3\times10^{5}~{\rm A}\cdot {\rm m}^{-1}$. Based on the same parameters and using $j_c\sim\bar j_c$, we obtain $\de V\sim 100~{\rm mV}$. Finally, from Eq.~\eqref{hint}, we have $\tau_{2q}/T_1\sim\tal\hbar\w_{10}/J'_{ij}$, where $\tau_{2q}$ is the two-qubit time scale. If we assume $J'_{ij}\sim J$, we obtain $\tau_{2q}/T_1\sim10^{-4}$. With $T_+\sim 3$~K, the neglect of the out-of-plane N\'eel vector oscillations [see discussion following Eq.~\eqref{afdyn}] is justified as long as the external field $B>2$~T. 

Qubit readout is achieved by bringing $\Delta U\rightarrow\hbar\omega_{10}$, which translates to $1-\eta\sim10^{-4}$ (using $B_{\rm ex}\sim100~\mbox{mT}$, $J_{\rm AF}/k_B\sim100~\mbox{K}$, and $L/a\sim125$) and $\Delta U\sim10^{-5}J$. Recall that $\Delta U\propto(1-\eta)^{5/4}$ and $\hbar\omega_{10}\propto(1-\eta)^{1/4}$. The difference in the exponents gives a separation of energy scales between the two variables so that, as $\eta\rightarrow1$, barrier $\Delta U$ can be lowered while keeping the qubit levels well-separated. Indeed, even for $1-\eta\sim10^{-4}$, $\hbar\omega_{10}$ remains of order $\sqrt{J/\chi}$.

We now estimate the dephasing rate Eq.~\eqref{dr}. Using $1-\eta\sim10^{-4}$ (together with the other parameters given above), we find 
\beq
\G_\varphi\sim(2.3\times10^{31}~\mbox{C}^{-2})S_I(0)\ .
\eeq
Modeling the current noise by a current source in parallel with a resistor $R$~\cite{martinisPRB03}, we may write $S_I(0)\sim 2k_BT/R$. At $T=1$~K, dephasing time of, e.g., $\tau_{\rm dep}=\G_\varphi^{-1}\sim1~\mu$s requires $R\sim1$~k$\W$.

The qubit's Rabi time is given by $t_0=\p/2\W_{\rm ac}$. To achieve $T_1/t_0\sim10^3$ (a thousand Rabi oscillations within the decoherence time defined by Gilbert damping), an ac charge current amplitude of $20~\mu$A would be necessary.

We now comment on the upper bound on temperature for the AF to operate in the mono-domain limit. Within the exchange approximation, a thermal magnon mode in the AF with wavenumber ${\protect {\boldsymbol q}}$ obeys a linear sound-like dispersion $\omega _{\protect {\boldsymbol q}}=v|{\protect {\boldsymbol q}}|$, where $v=\protect \sqrt {AL^3/\x}$ is the magnon velocity, and $A$ is the exchange stiffness of the AF. The temperature constraint for mono-domain operation then translates to $T\ll \hbar v\pi /k_BL\equiv T_0$. We then obtain $T_0\sim J_{\rm AF}(a/L)/k_B$. Estimating again for a cube-shaped KNiF$_3$ with $L=50$~nm, $J_{\rm AF}/k_B\sim100$~K and $L/a\sim125$, we obtain $T_0>T_+$. So for the relevant operational temperatures of the qubit, the mono-domain assumption is applicable.

To reduce the Joule heating caused by the current in the N, one may reduce $J$ in order to decrease $\bar j_c$. However, a decrease in $J$ will also lead to an undesirable decrease in $T_+$. Alternatively, $\bar j_c$ can be reduced with little effect on $T_+$ by increasing the effective spin Hall angle $\theta_{\rm SH}$; this may be achieved by using materials with strong effective spin Hall angles like topological insulators in place for the N. Achieving qubit operation at small currents is also important in order to minimize decoherence caused by shot noise in the current.

\section{Conclusions and general open problems}

In this work we introduce the first macroscopic spintronic qubit and discuss possible ways to make them interact quantum mechanically. In analogy with the corresponding superconducting devices including the current-biased Josephson qubits, superconducting quantum annealers and the SQUID magnetometers, our proposed device could, in principle, be used for preparing macroscopic quantum entanglement, probabilistic information processing, quantum annealing, and quantum-assisted sensing. Our estimation of the relevant physical
parameters based on the state-of-the-art technology shows qubit operational temperature that is more than an order of magnitude higher than the existing superconducting qubits, thus opening the possibility of macroscopic quantum information processing at temperatures above the dilution fridge range. While the relaxation time scale $T_1\sim10$~ns is relatively short compared to most superconducting devices, antiferromagnets are known to have internal magnetic dynamics in the THz range~\cite{kefferPR52,*gomonayLTP14}. Further research is
needed to demonstrate if such fast two-qubit time-scale gates could be experimentally achieved with our proposed AF qubits. 

In general, there are several engineering and technological challenges that have to be overcome in order to pave the way for potential practical relevance of any candidate solid-state quantum computing architecture. The nature of these obstacles have not yet fully understood even for superconducting devices that have been developed for several decades. For example, beyond the ratio of the Rabi frequency to decoherence rate, there are other systematic control errors that should be studied and improved, such as the origin of 1/f noise, cross-talks, and the limited bandwidth of control electronics leading to operating frequencies of about 10 GHz for existing superconducting qubits. Quantum hardware needs to be scaled up to a sufficient size to have any chance of being competitive with classical CMOS technologies that has been exponentially improved for over half of a century. Combining scaling and coherence is the big challenge of quantum systems engineering. This is fundamentally difficult due to non-separability of subsystems, and the no-cloning theorem, leading to design trade-offs that are global in nature. It is widely assumed that higher gate fidelities are the bottleneck for the scalability. This assumption leads to an oversimplification. The two-qubit fidelity is usually characterized in isolation, for optimized set of parameters, via randomized benchmarking. However, it should be noted that randomized benchmarking only gives a measure of the gate performance in the average case, but the threshold for fault-tolerant quantum computing is established for the worst-case gate performance which could lead to effective smaller tolerance to errors by one order of magnitude. Moreover, there are various mutli-qubit characterization of the proposed devices that has been traditionally overlooked. These multi-qubit performance issues are very important and are different in nature for the near-term digital shallow circuits and quantum processors. There are still several geometrical embedding limitations to achieve a desired computational complexity. These includes the degree of parallel operations, denser connectivity graphs, k-local interactions, maximum number of possible couplers, availability of the long-range interactions, and quantum state transfer limitations. There are other fundamental limits to the high precision tunability, fabrication, and control electronics scalability. These poses several fundamental physical and engineering trade-offs that are not yet fully understood for any of the major quantum computing proposals.

{\em Acknowledgments.}|S.~T. would like to thank Pramey Upadhyaya for discussions. This work was supported by FAME (an SRC STARnet center sponsored by MARCO and DARPA).


%

\end{document}